# Knowledge-Driven Mechanistic Enrichment of the Preeclampsia Ignorome

Tiffany J. Callahan[1,2†], Adrianne L. Stefanski[2], Jin-Dong Kim[3],
William A. Baumgartner Jr.[2], Jordan M. Wyrwa[4], Lawrence E. Hunter[2]

[1]*Department of Biomedical Informatics, Columbia University, New York, NY USA*
[2]*Computational Bioscience Program, University of Colorado Anschutz Medical Campus, Aurora, CO USA*
[3]*Database Center for Life Science, Research Organization of Information and Systems, Kashiwa, Japan*
[4]*Department of Physical Medicine and Rehabilitation, School of Medicine, University of Colorado Anschutz Medical Campus, Aurora, CO USA*
[†]*Email: tc3206@cumc.columbia.edu*

Preeclampsia is a leading cause of maternal and fetal morbidity and mortality. Currently, the only definitive treatment of preeclampsia is delivery of the placenta, which is central to the pathogenesis of the disease. Transcriptional profiling of human placenta from pregnancies complicated by preeclampsia has been extensively performed to identify differentially expressed genes (DEGs). The decisions to investigate DEGs experimentally are biased by many factors, causing many DEGs to remain uninvestigated. A set of DEGs which are associated with a disease experimentally, but which have no known association to the disease in the literature are known as the ignorome. Preeclampsia has an extensive body of scientific literature, a large pool of DEG data, and only one definitive treatment. Tools facilitating knowledge-based analyses, which are capable of combining disparate data from many sources in order to suggest underlying mechanisms of action, may be a valuable resource to support discovery and improve our understanding of this disease. In this work we demonstrate how a biomedical knowledge graph (KG) can be used to identify novel preeclampsia molecular mechanisms. Existing open source biomedical resources and publicly available high-throughput transcriptional profiling data were used to identify and annotate the function of currently uninvestigated preeclampsia-associated DEGs. Experimentally investigated genes associated with preeclampsia were identified from PubMed abstracts using text-mining methodologies. The relative complement of the text-mined- and meta-analysis-derived lists were identified as the uninvestigated preeclampsia-associated DEGs (n=445), i.e., the preeclampsia ignorome. Using the KG to investigate relevant DEGs revealed 53 novel clinically relevant and biologically actionable mechanistic associations.

## 1. Introduction

Preeclampsia has been known since Hippocrates described it in 400 BC and remains a leading cause of maternal and fetal morbidity and mortality.[1,2] Preeclampsia is a hypertensive, multisystemic disorder with an unknown etiology and variable maternal and fetal manifestations.[3] Maternally, preeclampsia presents as both hypertension and proteinuria, but can quickly progress to affect the kidneys, brain, and liver and in severe cases, results in thrombocytopenia, stroke, visual disturbance, renal failure, placental abruption, seizure, and death.[4] Fetal consequences of preeclampsia are a function of gestational age and the severity of the mother's condition, which may include intrauterine growth restriction (IUGR), prematurity, and perinatal death.[5]

Mechanistically, preeclampsia is thought to be partially caused by alterations in circulating angiogenic factors like vascular endothelial growth factor (VEGF), which is known to tightly regulate angiogenesis,[6] and triggers the development of organs. Preeclampsia is caused when free levels of transforming growth factor β (TGFβ), placental growth factor (PlGF), and VEGF are decreased, due to increased levels of antiangiogenic factors like soluble FMS-like tyrosine kinase 1 (Sflt-1) and Endoglin (sEng).[7] Despite extensive research and an in-depth understanding of the pathophysiology of preeclampsia, clinicians remain unable to prevent this disease.[8] One advantage of preeclampsia research is that upon termination of a pregnancy and/or delivery, the placenta is a non-vital organ and biopsies can be performed.[9] Even with this advantage and the sizable collection of transcriptomic data deposited in the public domain that has resulted from it, individual studies and many recent meta-analyses have not made much progress in furthering our understanding of effective prevention or treatment of preeclampsia.

In similarly complex diseases like asthma, strategies to identify relevant genes have yielded novel mechanistic insight into previously ignored genes.[10] The ignorome is defined as the portion of a gene signature shown to be significantly associated with a specific disease, but without a published mechanistic link — and often without any published disease association. Recently, researchers discovered that the top 5% of statistically significant differentially expressed genes (DEGs) were responsible for 70% of the published literature for a given disease.[11] Further examination of ignorome genes revealed no differences between the published and ignored genes in terms of their connectivity in co-expression networks; the biggest factor as to whether or not a gene was well-represented in the literature was its date of discovery.[11]

Preeclampsia has an extensive body of scientific literature, a large pool of DEG data, and only one definitive treatment. Given the rate at which science advances, tools facilitating knowledge-based analyses may be a valuable resource to support discovery and improve our understanding of this disease. Knowledge-based clinical research, and its ability to integrate disparate data from many sources in order to suggest underlying mechanisms of action, provides a potentially powerful new avenue to obtain mechanistic insight into experimental findings, such as in the enrichment of DEG lists. Very few DEGs are examined after an initial experiment because experimental follow-up is difficult and expensive, and nonsignificant DEGs are often investigated because prioritization approaches are generally based on experimental signal (e.g., effect size) rather than on existing knowledge. The goal of this paper was to demonstrate how a large-scale heterogeneous biomedical knowledge graph (KG) could be used to identify novel preeclampsia mechanisms from previously analyzed transcriptomic experiments.

## 2. Methods

The preeclampsia ignorome was identified in two steps: (i) identification of preeclampsia DEGs from multi-platform microarray meta-analysis and (ii) identification of genes associated with preeclampsia in the literature. The preeclampsia ignorome was generated from the set difference of the gene lists generated by these steps. Supplemental Material, code, and data are publicly available (http://tiffanycallahan.com/ignorenet/). Please see the analysis workflow readme (https://github.com/callahantiff/ignorenet/blob/master/analyses/preeclampsia/README.md) for information on the algorithms and data sources (KGs and gene lists) used for this analysis.

### 2.1. *Identification of the Preeclampsia Molecular Signature*

In collaboration with a PhD-level molecular biologist (ALS) who specializes in reproductive science, a meta-analysis was performed to identify relevant transcriptomic data on the Gene Expression Omnibus (GEO). Using the keyword "preeclampsia", publicly available human experiments deposited in GEO were examined. The initial set of identified studies were further reviewed for the following criteria to ensure: (i) processed samples were from a human placenta biopsy (i.e., chorionic villi, decidua basalis, and placenta); (ii) samples were processed using Agilent, Affymetrix, Applied Biosystems, Illumina, or NimbleGen; and (iii) studies provided normalized data and/or DEG lists. Each study's normalized data were processed using standard R pipelines using the ignorenet library (https://github.com/callahantiff/ignorenet). The final gene list was assembled by selecting significant DEGs ($p<0.05$) in at least 50% of the studies.

### 2.2. *Identification of Genes Associated with Preeclampsia in the Literature*

To identify known preeclampsia genes two strategies were employed: (i) **Literature-Driven.** This strategy aimed to identify relevant genes via keyword search against PubTator,[12] DisGeNET,[13] and Malacards (implemented 08-11/2017).[14] For this step, all queried results were manually verified for accuracy (i.e., verified that hits obtained were actually to preeclampsia and the associated keywords and were not errors or mismatches to closely associated synonyms or acronyms) and all valid associations were used to create a final unique list of genes; and (ii) **Gene-Driven.** This strategy aimed to identify relevant articles by querying 18 keywords in addition to the the preeclampsia molecular signature DEGs against PubAnnotation.[15] Similar to the Literature-Driven Approach, all results were manually verified for accuracy and all associations were used to create a final unique list of genes. See the Supplemental Material for keyword lists.

## 2.3. *Evaluation*

### 2.3.1. *Knowledge Graph Node Embeddings*

A v1.0 PheKnowLator KG[16] built using Linked Open Data and Open Biological and Biomedical Ontology Foundry ontologies was used for this analysis. The core set of ontologies included phenotypes (Human Phenotype Ontology [HP][17]), diseases (Human Disease Ontology [DOID][18]), and biological processes, molecular functions, and cellular components (Gene Ontology [GO][19]). Genes, pathways, and chemicals were added to the core set of ontologies to form the foundation of the KG which was extended by adding relations between phenotypes, diseases, and GO biological processes, molecular functions, and cellular components. Node embeddings were derived using C++ implementation of DeepWalk (hyperparameter settings suggested by developers: 512 dimensions, 100 walks, a walk length of 20, and a sliding window length of 10).[20]

### 2.3.2. *Visualizations*

Node embeddings were visualized using the t-distributed stochastic neighbor embedding (t-SNE) algorithm.[21] Experiments were performed to identify the best hyperparameter setting (perplexity=50). Node embeddings and ignorome genes were overlaid and visually inspected.

### 2.3.3. *Enrichment*

Using the node embeddings, the 100 nearest disease, drug, gene, GO concepts, pathway, and phenotype (i.e., domains) annotations for each ignorome gene as measured by pairwise cosine similarity (i.e., L2-normalized dot product of embedding vectors: $k(x, y) = \frac{xy^\top}{||x||\,||y||}$)[22] of the node embeddings were obtained. Annotations were reviewed by a PhD molecular biologist specializing in reproductive science (ALS; 08-09/2021). To determine if they occurred by chance, we:

1. Examined the overlap between the top-100 closest associations to each ignorome gene in the expert-verified list and the associations generated when enriching the preeclampsia ignorome using ToppGene;[23]
2. Computed how often the reviewed associations occurred by chance in 1,000 ignorome-sized random samples drawn from all non-ignorome genes represented in the KG. For each sample, the top-100 closest annotations to each gene, by domain were obtained and the number of annotations that overlapped with the expert-verified list was recorded. P-values were obtained for each domain by dividing the number of overlapping annotations out of the 1,000 samples, where a p-value of 0.05 indicates a 50 in 1,000 chance of observing a sample annotation that overlaps with the expert-verified annotations.

## 3. Results

### 3.1. *The Preeclampsia Ignorome*

As shown in Figure 1, there were 68 studies returned from the domain-expert review of GEO (Supplemental Table 1). Of these, 12 studies were determined to be eligible for inclusion in the current project (Supplemental Table 2). Processing these studies led to a sample of 548 DEGs, which appeared in 50% of the studies. The Gene-Driven strategy returned 1,962 articles which resulted in a total of 417 known preeclampsia genes. The Literature-Driven strategy returned 1,102 articles and 658 genes. These lists were combined and yielded a total of 946 unique genes associated with preeclampsia in the literature. Of the 548 genes identified as the preeclampsia molecular signature, 103 were found in the list of genes associated with preeclampsia in the literature, leaving 445 DEGs with no known literature evidence (i.e., "PE Ignorome" or non-overlapping blue circle of Figure 1). The remaining 843 genes associated with preeclampsia in the literature not found in the list of experimentally-derived genes are those that were found in less than 50% of studies, were not transcriptionally regulated, or played a role in the placenta.

The preeclampsia ignorome genes were examined for associations to other diseases in the literature. Figure 2, illustrates the number of articles from Malacards, DisGeNET, PubAnnotation, and PubTator that annotated each preeclampsia gene and the number of annotations to diseases other than preeclampsia that were found for each ignorome gene. Supplemental Table 3 contains the list of gene symbols binned by article count. As shown in Figure 2 (a), most genes were cited by fewer than 20 articles and less than 20 of the ignorome genes were cited more than 100 times. Among the genes cited 100 or more times were BRAF (n=2,749), TARDBP (n=694), and IDH1 (n=564). Figure 2 (b) illustrates the most frequently annotated diseases, which included neoplasms (n=1,778), mental disorders (n=280), and congenital diseases (n=272).

The PheKnowLator KG contained 128,286 nodes and 3,203,264 edges. The following 10 edge types, (ordered by frequency): drug-disease (n=1,216,900), drug-pathway (n=711,043), gene-gene (n=594,100), gene-go concept (n=265,002), gene-phenotype (n=120,288), gene-pathway (n=107,029), pathway-disease (n=106,727), disease-phenotype (n=43,817), gene-disease (n=20,452), and pathway-go concept (n=17,906), were used for the current analysis. The t-SNE plot is shown in Supplemental Figure 1 with nodes colored by node type and the preeclampsia genes marked using gold stars. As expected, most entities appeared closer to entities of a similar type than entities of other types except for GO concepts and phenotypes.

### 3.2. *Preeclampsia Ignorome Gene Enrichment*

Performing enrichment analysis on the preeclampsia ignorome genes using ToppGene returned 4,098 annotations (p<0.001 or Q-value Bonferroni <0.05). The annotations included four diseases, 3,667 drugs, 248 genes, 116 GO biological processes, 44 GO cellular components, 19 GO molecular functions, and no pathways or phenotypes. PheKnowLator node embeddings were used to annotate the preeclampsia ignorome genes by obtaining the 100 closest entities in vector space, which resulted in a total of 19 diseases (average similarity of 0.37 and frequency of 1.0 across the preeclampsia genes), 521 drugs (average similarity of 0.37 and frequency of 1.08 across the preeclampsia genes), 1,060 GO concepts (average similarity of 0.38 and frequency of 1.49 across the preeclampsia genes), 563 pathways (average similarity of 0.44 and frequency of 2.29 across the preeclampsia genes), and 64 phenotypes (average similarity of 0.30 and frequency of 1.0 across the preeclampsia genes). None of the identified diseases, GO concepts, pathways, or phenotypes overlapped with the ToppGene annotations, but seven of the identified drugs and 188 of the identified genes did.

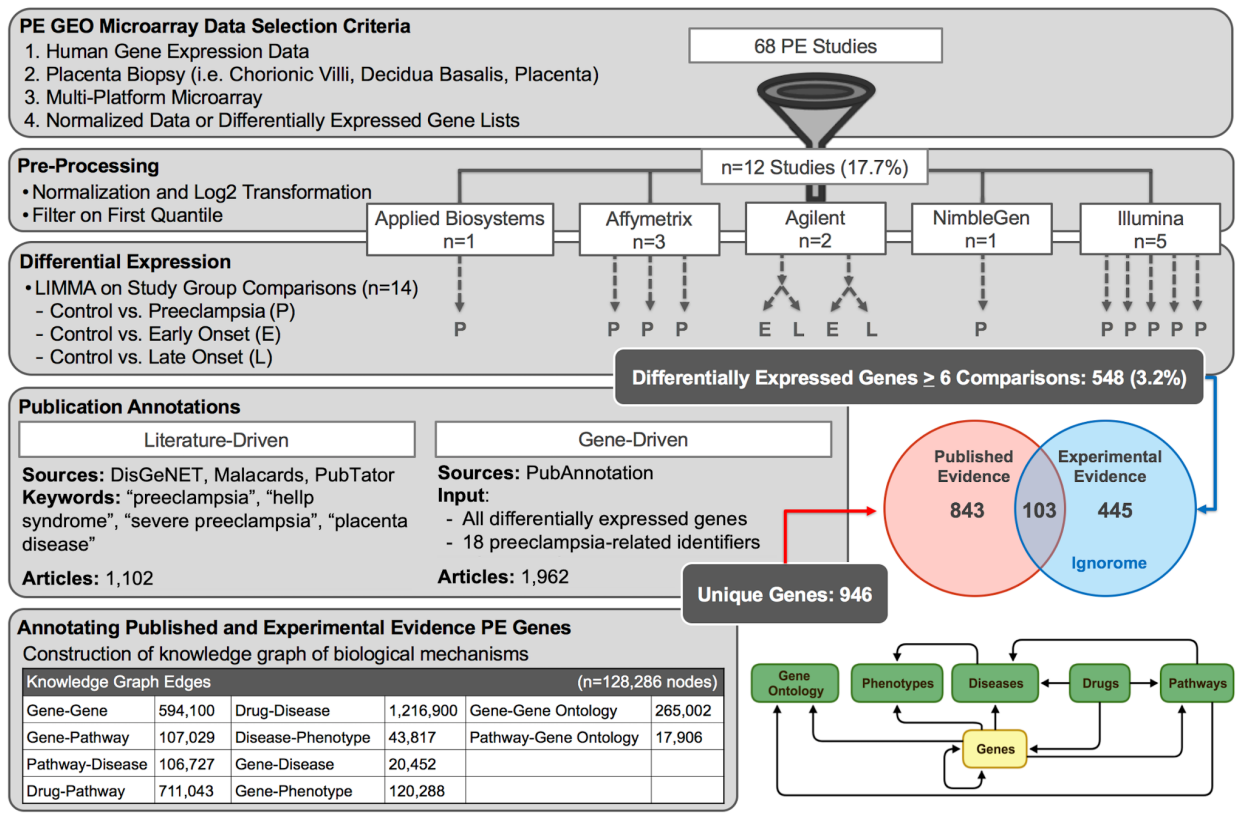

Fig. 1. Overview of Results for Finding the Preeclampsia Ignorome. The figure provides an overview of the procedures utilized in order to obtain the preeclampsia ignorome. Acronyms - PE: Preeclampsia.

The reproductive science expert reviewed the KG-derived annotations and provided explanations using her domain expertise and rigorous literature review, which resulted in the validation of 53 annotations and included five phenotypes (Supplemental Table 4), 10 pathways (Supplemental Table 5), 10 drugs (Supplemental Table 6), 10 genes (Supplemental Table 7), 10 GO concepts (Supplemental Table 8), and eight diseases (Supplemental Table 9). The expert spent

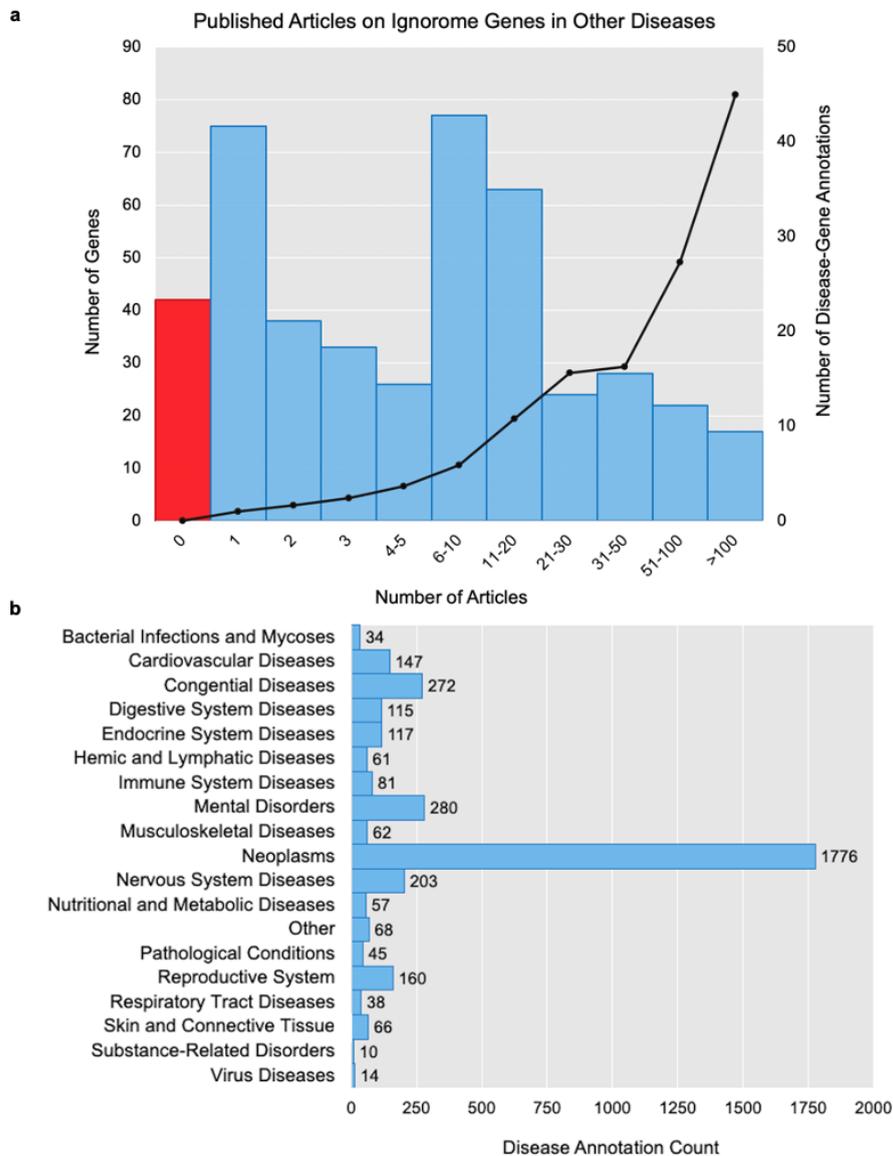

Fig. 2. Preeclampsia Ignorome Gene Annotations in Other Diseases. (a) illustrates the literature coverage of the 445 preeclampsia ignorome genes to other diseases. The x-axis represents the number of disease-annotated articles for each gene. The left y-axis shows the number of genes as bars, where the red bar contains the number of genes with no literature annotations to any disease. The right y-axis shows the number of diseases annotated to each preeclampsia gene and the number of annotations to diseases other than preeclampsia that were found for each ignorome gene in the literature. (b) Plots the counts of literature annotations to high-level disease categories.

~six hours on this task, noting that the drug and disease associations were the most challenging and time consuming to review. For all tables, evidence is provided in the form of mechanistic explanations and includes support from peer reviewed articles. None of the expert-reviewed annotations occurred by chance (*ps<0.005*): (i) **Diseases.** 485 concepts with an average similarity of 0.40 (0.26-0.77); (ii) **Drugs.** 8,371 concepts with an average similarity of 0.41 (0.25-0.69); (iii) **Genes.** 23,728 concepts with an average similarity of 0.47 (0.24-0.93); (iv) **GO Concepts.** 15,447 concepts with an average similarity of 0.39 (0.25-0.77), four overlapped with ToppGene (i.e., GO:0000398, GO:0005747, GO:0070125, and GO:0005833); (v) **Pathways.** 1,671 concepts with an average similarity of 0.45 (0.24-0.77), four overlapped with ToppGene (i.e., R-HSA-194840, R-HSA-611105, R-HSA-5419276, and R-HSA-6799198]); and (vi) **Phenotypes.** 3,080 concepts with an average similarity of 0.36 (0.25-0.63), one overlapped with ToppGene (i.e., HP:0008316).

## 4. Discussion

Recent examination of the ignorome genes has revealed an interesting phenomena; the only difference between the genes that are frequently published for a given disease and those that are not is the date in which the genes were discovered.[11] This presents new exciting opportunities for discovery, especially with respect to improving our understanding of complex diseases like preeclampsia. Given the rate at which science advances and the volume of data that is generated as a result, tools facilitating knowledge-based analyses are valuable resources to support discovery. This paper demonstrates how a large-scale biomedical KG could be used to identify novel clinically relevant and biologically actionable preeclampsia mechanisms from previously analyzed experiments. Although limited, similar work has demonstrated the value of using KGs to generate new disease-associated genes,[25,26] drug-target interactions,[27] and evaluate the consistency of genome annotations through biological pathways.[28] A big difference between these methods and ours is the depth and breadth of knowledge covered by our KG and that we are able to generate explanations that consist of multiple types of biological entities. To the best of our knowledge, our work is the first to perform KG-based mechanistic enrichment of the preeclampsia ignorome.

### 4.1. *Novel Preeclampsia-Associated Mechanisms*

Precise characterization of phenotypes will require the ability to identify and understand complicated biological relationships. Our novel preeclampsia ignorome associations required fairly complicated explanations. A few relevant results from each domain are described below.

**Phenotypes.** These associations present new opportunities to enrich our understanding of the phenotypic variance within preeclampsia. There were many interesting associations, but one of the most relevant was PPM1K to *Elevated Plasma Branched Chain Amino Acids*. Examining this mechanism closer revealed that the disruption of PPM1K results in an increase of branched chain amino acids, which can result in oxidative stress, insulin resistance, and eventually obesity, by activation of the mammalian target of rapamycin complex 1 (mTORC1) signaling.[29] mTORC1 signaling is vital for communicating placental growth factor signaling and when reduced in IUGR pregnancies, has been found to impair mitochondrial respiration and lead to placental insufficiency.[30] While mitochondrial dysfunction is known to be central to preeclampsia pathophysiology,[31] the role of PPM1K in preeclampsia has yet to be thoroughly examined.

**Pathways.** Associations within this domain highlight potential new avenues of investigation for specific gene targets within pathways that are known to play a role in preeclampsia. Three associations are highlighted: (i) MFAP5 and FBLN5 to the *Elastic Fibre Formation pathway* – this pathway is altered in umbilical cord vessels from pregnancies complicated by preeclampsia,[32] but the exact molecular mechanism causing the alteration is unknown; (ii) ADAMTSL3 and SPON1 to *Diseases Associated with O-glycosylation of Proteins* – it is known that altered o-glycosylation is associated with aberrant immune cell dynamics at the maternal-fetal interface[33] and in severe preeclampsia, altered glycosylation of maternal plasma proteins is associated with increased monocyte adhesion;[34] and (iii) TCP1, RGS11, and TBCD to *Protein Folding*; the impact of aberrant protein folding on preeclampsia is well documented[35] but the roles of TCP1, RGS11, and TBCD in this pathway are not fully understood.

**Drugs.** The association of MME to *anti-asthmatic agents* may provide an avenue for drug repurposing. Membrane matrix remodeling is critical to placental development[36] and women who experience asthma during pregnancy have an increased risk of developing preeclampsia.[37] While beta-adrenergic agonists such as ritodrine and terbutaline have been used for the management of asthma and preterm labor, it is unclear as to whether or not anti-asthmatic medications could reduce the risk of preeclampsia.[38]

**Genes.** Associations within this domain may provide a deeper understanding of the molecular landscape of preeclampsia by helping researchers identify relevant, yet understudied genes, for example, the associations from PLOD1, FBLN5, and PTGDS to PLOD2. These associations are supported by evidence that PLOD2 is a protein that is upregulated in trophoblast stem cells cultured under hypoxic conditions.[39]

**GO Concepts.** These associations may highlight opportunities to bridge findings across domains, for example, the associations between ACTR3, NEBL, ACTR3B, MYO1B, COBLL1, ZNF185, and ITPRID2 to the GO Molecular Function *Actin Filament Binding*. Preeclampsia is associated with altered actin polymerization via endothelial protein C receptor.[40] Traditionally, actin has been studied via cell biology or histology but a deeper examination of these associations within the biological context of preeclampsia has the potential to connect the findings derived from these disconnected studies.

**Diseases.** By enriching microarray data derived from placental samples with KG-based mechanisms it is possible to identify diseases that occur later in life, but which are likely to be associated with fetal exposure to maternal preeclampsia. For example, the association between STS and *Attention Deficit Hyperactivity Disorder* (ADHD); STS dysfunction causes ADHD[41] and offspring of preeclamptic mothers[41] are more likely to be diagnosed with ADHD.[42]

### 4.2. *Preeclampsia Ignorome Enrichment*

Examining differences in the enrichment of GO annotations relevant to preeclampsia revealed some interesting insights. For example, *Placenta Development* included 25 genes associated with preeclampsia in the literature, 10 genes with both literature and experimental evidence, but none were ignorome genes. This finding confirms our expectations – a lot of genes known to impact placental development exist and many have been investigated experimentally. In contrast, the *Cell Surface Receptor Signaling Pathway* included genes from all three of the aforementioned groups, supporting our observation that the things enriched for this biological process are over-studied. Only ~10% of the ignorome genes (n=42) had no other disease annotations when examining the coverage of ignorome genes in the literature. This leaves a significant body of literature spanning a wide-range of diseases, which would take a substantial amount of time and domain expertise, a task which is often out-of-scope for most researchers.

### 4.3. *Limitations and Future Work*

Our work has important limitations: (i) all analyses were performed using data available in 2017. More data has likely become available since then, but re-analysis of these data was not feasible; (ii) microarray data were only obtained from GEO. It is important to explore other repositories and other types of molecular data; (iii) the pipeline depends on tools like PubTator to review the literature and domain experts to formulate explanations for annotation. Incorporation of more advanced models and pipelines would improve scalability and reduce bias; (iv) our results require additional validation (i.e., wet lab and sensitivity analysis/ablation studies) before the full utility of our approach can be determined; and (v) the PheKnowLator Ecosystem is new and while preliminary studies have suggested it produces robust KGs additional experiments are warranted. Future work aims to address these limitations and will explore advanced algorithms to process novel associations like natural language generators.

## 5. Conclusion

Large-scale biomedical KGs new opportunities to improve our understanding of complex diseases, like preeclampsia. With assistance from a domain expert, we propose potential mechanistic explanations for 53 new associations between preeclampsia ignorome genes. These mechanistic explanations represent biologically-actionable discoveries that await further investigation in the hopes of finding a means to prevent preeclampsia.

**Acknowledgements**

This work was supported by the National Library of Medicine (T15LM009451).